\documentstyle[psfig,aaspp4,11pt]{article}
\def\scrL{{\mathcal L}}
\begin{document}
\title{X-ray Signatures of Evolving Radio Galaxies} \author
{Sebastian~Heinz\altaffilmark{1,4},
Christopher~S.~Reynolds\altaffilmark{2}, \and
Mitchell~C.~Begelman\altaffilmark{3,4}} \affil{JILA, University of Colorado
and National Institute of Standards and Technology, Boulder, Colorado
80309-0440} \altaffiltext{1}{email address: heinzs@bogart.Colorado.edu}
\altaffiltext{2}{email address: chris@rocinante.Colorado.edu}
\altaffiltext{3}{email address: mitch@jila.Colorado.edu} \altaffiltext{4}{
also at Department of Astrophysical and Planetary Sciences, University of
Colorado, Boulder}

\begin{abstract}
We present a simple model for an evolving  radio galaxy, as it expands into
the hot,  X--ray emitting interstellar medium (ISM)  of its host  galaxy or
similarly into  the hot intracluster medium  (ICM) of its host  cluster. We
solve the governing equations numerically on a  grid of model parameters in
order to present simple analytical   tools for X--ray observations of   the
shocked shell that is pushed out by the cocoon and the associated cavity in
the  cluster emission. We apply  these tools to the   well known example of
Perseus A to  show  that  its  time-averaged kinetic luminosity    probably
exceeds  $10^{46}\,{\rm  ergs\,sec^{-1}}$, much  larger than  the estimated
current  power.  We  show how future  observations  can be  used to extract
useful source  parameters such as the average  kinetic power and the source
age, and discuss detectability of sources at various stages of their lives.
\end{abstract}

\section{Introduction}
In recent  years there has been much  progress in understanding how various
classes of  powerful  extragalactic radio  sources can  be described in the
context  of an evolutionary  picture.  Recent radio surveys have identified
classes  of powerful  sources which are  morphologically  similar to  FR II
radio  galaxies but appreciably   smaller.   Sources less than 500\,pc   in
extent have  been termed  Compact   Symmetric Objects  (CSOs; Wilkinson  et
al.~1994), whereas those in the  size range 0.5--15\,kpc are often referred
to as Medium Symmetric Objects (MSOs; Fanti et al.~1995).  These classes of
small   sources,   which  form    approximately  one   quarter  of  current
flux--limited radio surveys, are thought  to correspond to the early stages
of   full-sized FR-II radio  galaxies  (Begelman 1996; Readhead, Taylor, \&
Pearson 1996).

Central to our understanding of these sources is the following theoretical
picture (first proposed by Scheuer 1974).  Relativistic plasma flows from
the central AGN in the form of collimated jets, passes through terminal
shocks corresponding to the radio hot-spots, and inflates a `cocoon' which
envelops the whole source.  This cocoon becomes highly overpressured with
respect to the surrounding interstellar/intracluster medium and, hence,
drives a strong shock into this material.  The swept-up material forms a
dense shell separated from the cocoon by a contact discontinuity (see,
e.g., Begelman \& Cioffi 1989).  In the late stages of evolution, the
expansion of the cocoon/shell becomes subsonic and the cocoon disrupts and
mixes with the ambient medium. 

Although low-frequency radio observations do reveal well-defined
synchrotron emitting cocoons (e.g., Cygnus-A; Carilli, Perley, \& Harris
1994), there is relatively little direct observational evidence for the
shocked shell.  In principle, there are at least two methods of detecting
the shell.  Firstly, one can search for the optical line emission that is
excited near the shock front (see, e.g., Bicknell \& Begelman 1996 for an
explanation of the H$\alpha$ line emission in M87).  Such line emission is
very sensitive to unknown parameters such as the fraction of cold material
in the surrounding ISM/ICM and the ionization state of that material.
Secondly, one can search for the X-ray emission from the shocked ISM/ICM
and the associated cavity in the ambient material.  Cavities in the ICM
have been observed in Cygnus A (Carilli et al.~1994) and Perseus A
(B\"ohringer et al.~1993). 

In this paper we will discuss these X-ray signatures of expanding powerful
radio sources.  We will show how X-ray imaging observations can be used to
constrain important source parameters such as the mean kinetic energy
output of the object and its age, which are otherwise difficult to
constrain directly.  We apply our method to current {\em ROSAT} HRI data on
Perseus A and suggest that this classic source is in fact in a very
quiescent state.  Furthermore, we develop diagnostics that can be used to
interpret the data from the new generation of high resolution X-ray
imagers, starting with {\em AXAF} in 1998. 

In \S \ref{sec:model}, we describe the model and the basic assumptions that
we use for our study.  In \S \ref{sec:examples}, we apply this model to
Perseus A, and in \S \ref{sec:predictions}, we present a set of diagnostic
tools that can be applied to {\em AXAF} data.  In \S \ref{sec:discussion},
we discuss further aspects of this study and summarize the important
points.

\section{Description of the Model}
\label{sec:model}
\subsection{The Dynamical Model}
We first outline the basic assumptions used in our model to find a simple,
robust description of the early stages of radio galaxy evolution into a
surrounding hot medium.  Our model is based on the analysis by Reynolds \&
Begelman (1997).  Following this work, we make several simplifying
assumptions: 
\begin{enumerate}
\item{Spherical symmetry. For the purpose of this paper it is sufficient to
neglect the prolate structure observed in most radio sources, since more
detailed hydrodynamics would be required in order to determine the shape of
the cocoon beyond a self--similar form (see, e.g., Clarke, Harris, \&
Carilli 1997).  The level of detail required in such simulations and the
amount of computing power necessary to explore parameter space in the
desired manner would defeat the scope of this paper.  The observed
elongations are moderate (axial ratios of order 3 in FR II sources, see
Carilli et al.~1994).  The dependence of our results on the source radius
is relatively weak, which makes us confident that the application of our
model to non-spherical sources will introduce minor errors only.} 
\item{Purely relativistic gas inside the cocoon (i.e., the adiabatic
index in the cocoon is $\gamma_{\rm c}=\frac{4}{3}$) and non-relativistic
gas in the swept up shell ($\gamma_{\rm s}=\frac{5}{3}$).  The latter
assumption is valid in all but the early stages of the most luminous
sources, in which the electrons become relativistic.} 
\item{Uniform pressure.  We take the pressure in cocoon and shell to be
uniform and equal.  This is a reasonable approximation in the context of
the previous assumption, as the sound speed inside the cocoon
($c/\sqrt{3}$) will be significantly higher than the expansion velocity of
the shell.  The radio hot spots will be overpressured, but we will neglect
this complication in the following.} 
\item{A King-model X--ray atmosphere provided by either the host galaxy or
the cluster in which the AGN is embedded. The density profile thus behaves
as: 
\begin{equation}
\label{eq:density}
\rho(r)=\rho_{\rm 0}\left[ 1 + \left(\frac{r}{r_{\rm
c}}\right)^{2}\right]^{-\frac{3\beta}{2}}, 
\end{equation}
where $\rho_{\rm 0}$ is the central density and $r_{\rm c}$ is the core
radius.  $\beta$ can take any positive value; however, for the interface
between cocoon and shell to be stable against Rayleigh--Taylor instability
we need to assume $\beta\lesssim\frac{2}{3}$ (see below).  $\beta$ is
observationally determined by the ratio of the velocity dispersion of the
cluster galaxies to the temperature of the cluster gas.} 
\item{Non-radiative shocks. We neglect energy loss due to radiative losses
in the equations below. This is justified as long as the cooling time is
long compared to the source lifetime, a condition satisfied in the
parameter range we consider.} 
\end{enumerate}

The system is well defined by energy conservation within the cocoon and the
shell: 
\begin{equation}
\label{eq:cocoon}
\frac{1}{\gamma_{\rm c}-1}\left(V_{\rm c}\dot{p}+\gamma _{\rm c}\dot{V_{\rm
c}} p\right)=L(t) 
\end{equation}
and 
\begin{equation}
\label{eq:shock}
\frac{1}{\gamma_{\rm s}-1} \left( V_{\rm s}\dot{p} + \gamma _{\rm
s}\dot{V_{\rm s}} p \right) = \frac{4 \pi }{2} r_{\rm s}^{2} \rho (r_{\rm
s}) \dot{r_{\rm s}}^{3}, 
\end{equation}
and the ram pressure condition at the shock: 
\begin{equation}
\label{eq:pressure}
p_{\rm c,s}(t)=\rho(r_{\rm s}){\dot{r_{\rm s}}}^{2}. 
\end{equation}
Here, $V_{\rm s}$ and $V_{\rm c}$ are the shell and cocoon volumes,
respectively, $r_{\rm s}$ is the shock radius, and $p_{\rm c,s}=p(t)$ is
the (uniform) interior pressure, which is a function of time.  $L(t)$ is
the kinetic luminosity of the jets feeding the cocoon.  A dot indicates a
time derivative, i.e., $\dot p=dp/dt$.  Equation (\ref{eq:pressure})
holds only in the case of supersonic expansion, a condition well satisfied
in the early evolutionary stages of our models but which is violated as
sources pass a characteristic size.  Once a source has decelerated below
the ambient sound speed (typically of the order of $c_{\rm sound}\lesssim
1000{\rm\,km\,sec^{-1}}$) the evolution will resemble an expansion wave
rather than a shock wave.  The shell will thin out and eventually blend
into the ambient medium; the cocoon--shell interface will become unstable
and collapse on timescales of order the free fall time.  The kinetic
luminosity $L(t)$ in equation (\ref{eq:cocoon}) can in general be time
dependent to allow for the intermittency suggested by Reynolds \& Begelman
(1997, see \S \ref{sec:intermittency}).  For now, we will take it to be
constant.  In a sense, this can be interpreted as a time averaged
luminosity $\scrL\mathit \equiv \langle L(t)\rangle$. 

To explore parameter space we have integrated equations (\ref{eq:density})
to (\ref{eq:pressure}) numerically over a time span of $10^8$ years,
assuming an initially small source.  Our models were calculated over a grid
of input parameters $\scrL$ and $r_{\rm c}$. We used the following
parameter values: 
\begin{itemize}
\label{items}
\item{Luminosities ranging from $\scrL = 10^{42}$ to
$10^{52} {\rm\,ergs\,sec^{-1}}$.  We use a fiducial value of ${\mathcal
L}=10^{46}{\rm\,ergs\,sec^{-1}}$ throughout the paper except where
indicated. Note that luminosities in excess of $\scrL \approx 10^{48}\,{\rm
ergs\,sec^{-1}}$ can be considered unphysical, since they correspond to
Eddington luminosities for black hole masses $\gtrsim 10^{10}\,M_{\odot}$.
However, as will be shown below, only the combination $\scrL/\rho_{\rm 0}$
is relevant to the dynamics, so we chose to hold $\rho_{\rm 0}$ fixed and
explore a wide range of $\mathcal L$. Unphysically high values of $\scrL$
can be interpreted as relevant to low density sources.} 
\item{Core radii in the range of $50\,{\rm pc}\leq r_{\rm c}\leq 500\,{\rm
kpc}$, with a fiducial value of $500$\,pc, typical of elliptical galaxies.} 
\item{We set $\beta\equiv\frac{1}{2}$ throughout the rest of the paper,
corresponding to $\rho \propto r^{-1.5}$ for $r \gg r_{\rm c}$.} 
\item{We fixed the central density to be $\rho_{\rm 0}=1.7\times
10^{-25}\,{\rm g\,cm^{-3}}$ or $n_{\rm e,0}= 0.1\,{\rm cm^{-3}}$}. 
\end{itemize}

For power-law density distributions $\rho\propto r^{-\alpha}$ and constant,
non--zero kinetic luminosity, a self-similar solution to the equations is
possible.  This solution is a good indicator of how the more general
solution scales with the input quantities $\mathcal L$, $\rho_{\rm 0}$, and
$r_{\rm c}$.  A necessary condition for a self-similar solution is that the
cocoon radius $r_{\rm c}$ be a fixed fraction $\lambda$ of the shell
radius, i.e., $r_{\rm c}\equiv \lambda r_{\rm s}$.  Under these
assumptions, equations (\ref{eq:cocoon}) to (\ref{eq:pressure}) yield 
\begin{equation}
\label{eq:lambda}
\lambda=\left(\frac{9\gamma_{\rm s}-2\alpha +1}{18\gamma_{\rm s}-2\alpha
-8}\right)^{\frac{1}{3}} 
\end{equation}
\begin{equation}
\label{eq:selfsimilar}
r_{\rm s}(t)=r_{\rm 0}\,\left(\frac{t}{t_{\rm
0}}\right)^{\frac{3}{5-\alpha}}, 
\end{equation}
where $r_{\rm 0}$ is the shell radius at time $t_{\rm 0}$, which is defined
by 
\begin{equation}
\label{eq:scaling}
t_{\rm 0}=C_1\times\left(\frac{\rho_{\rm 0}\,{r_{\rm 0}}^5}
\scrL\right)^{\frac{1}{3}} 
\end{equation}
with 
\begin{equation}
\label{eq:scaleconstant}
C_1\equiv\left\{\frac{36\pi \lambda^{3}}{\left(5-\alpha\right)^{3}
\left(\gamma_{\rm c}-1\right)}\left[3\gamma_{\rm c}-1+\frac{2}{3}
\left(\alpha-2\right)\right]\right\}^{\frac{1}{3}} 
\end{equation}

It follows from equation (\ref{eq:pressure}) and the assumed pressure
equilibrium between cocoon and shell that the shell temperature (assuming
an ideal, non--relativistic gas) is proportional to the square of the
expansion velocity, ${\dot{r_{\rm s}}}^2$.  Thus, for a given $r$ the
temperature in the self-similar solution goes as 
\begin{equation}
\label{eq:Tscaling}
T\propto {\dot{r_{\rm s}}}^2 \propto 1/{t_0}^2 \propto
\left(\scrL/\rho_0\right)^{2/3}. 
\end{equation}

Equation (\ref{eq:selfsimilar}) reveals that solutions for $\alpha \geq 2$
(i.e., $\beta \geq \frac{2}{3}$ in the limit $r \gg r_{\rm c}$) are
Rayleigh--Taylor unstable, since for those values the cocoon--shell
interface is always accelerated and the shell is very dense compared to the
cocoon gas. 

We can see from the basic set of equations that the only two parameters
entering the solution are a radial scale factor ($r_{\rm c}$ in the
King-profile case, $r_{\rm 0}$ in the self-similar case) and
$\scrL/\rho_{\rm 0}$ --- this statement holds even for arbitrary density
profiles.  While $r_{\rm c}$ and $\rho_{\rm 0}$ can in principle be
determined by direct observation, $\scrL$ can only be inferred
theoretically from observed radio brightnesses.  However, this conversion
is not trivial and it would be very useful to constrain the kinetic
luminosity directly. 

\subsection{Calculation of the X--Ray Brightness}

Because equations (\ref{eq:density}) to (\ref{eq:pressure}) do not specify
the density and temperature structure inside the shell itself, we have to
make additional assumptions about the radial dependences of $\rho$ and $T$.
In this paper we take both variables to be uniform within the shell and use
the ideal gas law and mass conservation of the swept up material to convert
from the pressure given by equation (\ref{eq:pressure}) to the temperature
$T_{\rm shell}$.  The shock jump conditions dictate the values of
$\rho_{\rm shock}$ and $T_{\rm shock}$ immediately behind the shock.  In
the case of a strong shock, the density jumps to $\rho_{\rm shock}\sim
4\rho_{\rm preshock}$, and the temperature jumps to $kT_{\rm shock}\sim
2\frac{\gamma_{\rm s}-1}{\left(\gamma_{\rm s}+1\right)^{2}}\ \mu
\dot{r_{\rm s}}^{2}$ where $\mu$ is the molecular weight.  Comparing
$\rho_{\rm shock}$ and $T_{\rm shock}$ to the average shell values reveals
that the shell is overdense in most cases, i.e., on average the shell
is colder than the most recently shocked material, as is expected due to
adiabatic expansion.  The density ratio is $\approx 6$ over most of the
parameter range, close enough to the strong shock jump value of 4 to assume
near uniformity of $\rho$ within the shell. 

We calculate the specific X--ray emissivity due to thermal bremsstrahlung,
taking the material to be composed of fully ionized hydrogen only (Rybicki
\& Lightman 1979): 
\begin{equation}
\label{eq:brems}
\frac{dW}{dVdt\ d\nu}=\frac{2^5\pi e^6}{3 m c^3}\left(\frac{2 \pi}{3 k
m}\right)^{1/2}T^{-1/2}n_{\rm e}^{2}e^{-h\nu/kT}\bar{g}_{\rm ff}. 
\end{equation}
Since we are only interested in supersonic cases, we will have to discard
parameter values for which the temperature drops below the ambient
temperature $T_{\rm ambient}\approx 1-10$\,keV at a given radius.  Because
all high resolution imaging X-ray facilities available in the near future
have bandwidths not far above this range, we can assume $kT > h\nu \gg
13.6$\,eV, thus the Gaunt factor is given by the small angle uncertainty
principle approximation: 
\begin{equation}
\label{eq:gaunt}
\bar{g}_{\rm ff}=\frac{\sqrt{3}}{\pi}\ln{\left(\frac{4}{\zeta}\frac{kT}
{h\nu}\right)}, 
\end{equation}
where $\zeta=1.78$ is Gauss's number. 

We also use equations (\ref{eq:brems}) and (\ref{eq:gaunt}) to compute the
thermal bremsstrahlung emission from the hot galaxy/cluster gas, which we
assume to be {\em isothermal} throughout the rest of the paper.  Clearly,
the existence of a massive cold component to the ISM/ICM (as expected if a
cooling flow operates) will alter the properties of the solution, in
particular the emissivities, as relatively cold, dense material will
radiate more efficiently.  This aspect will be commented on in \S
\ref{sec:cold}. 

\subsection{Observational Diagnostics}
\label{sec:diagnostic}

\begin{figure}
\centerline{
\psfig{figure=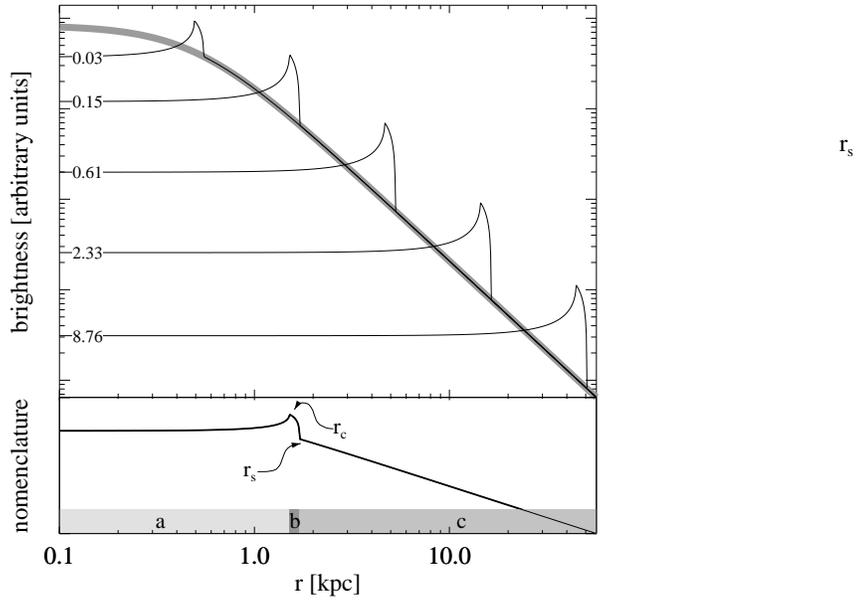,width=.5\columnwidth}}
\caption{Source evolution, seen through a flat $1-5$\,keV
bandpass for our fiducial parameters (see \S \ref{sec:model}).  The x-axis
is in units of kpc, the y-axis in arbitrary flux units.  The different
curves correspond to radial profiles at different times as labeled in the
figure (in units of $10^{6}$\,years). The thick grey curve corresponds to
$t=0$, i.e., the undisturbed cluster profile.
\label{fig:timesequence}}
\end{figure}

Figure\,\ref{fig:timesequence} shows the surface brightness results ({\em
i.e.}, the emissivity integrated along the line of sight) of integrating
the model with our fiducial parameters.  Shown are radial profiles at
different times as indicated in the figure.  Three basic features are
identifiable from the figure, indicated by shadowed regions on the bottom
of the plot: 
\begin{itemize}
\item[a)]{The flat part inside the shell, steepening into the bright
shell. This component includes all lines of sight penetrating the cocoon,
i.e., $r\,<\,r_{\rm c}$.} 
\item[b)]{The shell. We define this part as all lines of sight outside the
cocoon but still penetrating the outer shell, i.e., $r_{\rm
c}\,<\,r\,<\,r_{\rm s}$.} 
\item[c)]{The undisturbed cluster emission, i.e., $r\,>\,r_{\rm
s}$. This part simply tracks the King-profile atmosphere. In our case,
since we used an index of $\beta=\frac{1}{2}$, the surface brightness in
the power law part goes as $r^{-2}$ and flattens into the core.} 
\end{itemize}

We identify the following readily--measurable diagnostics which will be
used in the rest of this paper to investigate source parameters: 
\begin{itemize}
\item[i)]{{\em shell-cluster ratio}: the ratio of the surface brightness at
the line of sight tangential to the cocoon (at $r_{\rm c}$) to the surface
brightness at the line of sight tangential to the shell (at $r_{\rm s}$).
The emission from the shell (without any contribution from the X--ray
atmosphere) is brightest along the former line of sight.  The latter is the
brightest line of sight outside the region of shell emission.  A high
contrast is important for the detectability of the source.  Since the
dynamical solution depends only on $\scrL/\rho_{\rm 0}$, we can see that
brightness ratios are also going to depend on $\scrL$ and $\rho_{\rm 0}$
only in the combination $\scrL/\rho_{\rm 0}$, since the density
normalization $\rho_{\rm 0}$ cancels from equation (\ref{eq:brems}).  For a
purely self-similar solution, we can thus expect this ratio to be
proportional to $T^{-1/2}\propto (\scrL/\rho_{\rm 0})^{-1/3}$ [see equation
(\ref{eq:Tscaling})].}
\item[ii)]{{\em center-cluster ratio}: the ratio of the surface brightness
at the central line of sight to the surface brightness at the line of sight
tangential to the shell. This ratio indicates if the central lines of sight
are brightness enhanced or depressed compared to the cluster emission, {\em
i.e.}, if there is an `X--ray hole'.  Again, since it is a brightness
ratio, the {\em center-cluster ratio} should depend only on
$\scrL/\rho_{\rm 0}$ and $r_{\rm c}$, and for the self-similar case it
should go as $T^{-1/2}\propto (\scrL/\rho_{\rm 0})^{-1/3}$.  The presence
of a strong point-like AGN X-ray component will swamp the cluster emission
at the very center.  Because the brightness profile is very flat at central
lines of sight (as can be seen from Fig.\,\ref{fig:timesequence}), we can
avoid the contamination by taking an off-center value for the central
surface brightness and will only make a small error.} 
\item[iii)]{{\em shell count rate $\zeta$}: the integrated count rate from
all lines of sight penetrating the shell (in other words: all of areas {\em
a} and {\em b} in Fig.\,\ref{fig:timesequence}), including back-- and
foreground emission from the X--ray atmosphere. This quantity is easier to
determine than the background subtracted emission from just the
shell. Notice, however, that since it depends on absolute normalization,
both the distance to the object $d$ and the density normalization
$\rho_{\rm 0}$ factor into the {\em shell count rate $\zeta$}, thus we
cannot express it as a function of $\scrL/\rho_{\rm 0}$ and $r_{\rm c}$
only, rather, a factor of $(\rho_{\rm 0}/d)^2$ remains.} 
\end{itemize}
The particular values of these diagnostics for a given source will depend
on the assumed instrumental response, as the cluster and shell have
different temperatures and thus different spectra.  In the following we
will use both the {\em ROSAT} HRI band and the predicted {\em AXAF} ACIS
band as indicated. 

\section{Applications to Existing Data}
\label{sec:examples}
\subsection{Perseus A}
\label{sec:PerA}

\begin{figure}
\centerline{
  \psfig{figure=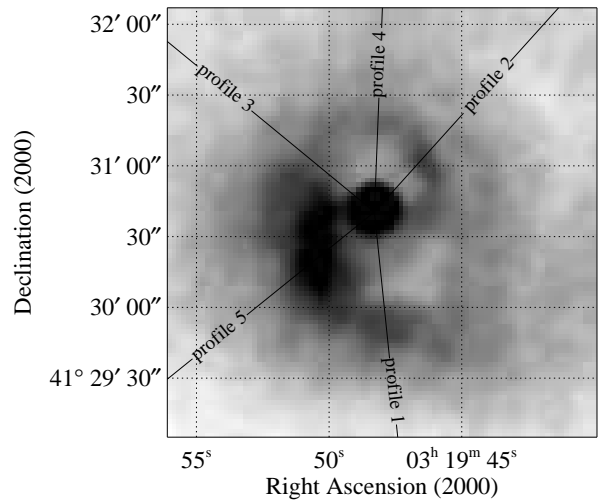,width=.5\columnwidth}}
\caption{A 50\,ksec {\em ROSAT} HRI exposure of Per A, smoothed
with a 2 arcsec Gaussian beam.  We have chosen the contrast to emphasize
the shell structure.  These data were downloaded from the LEGACY public
archive at GSFC (NASA).  The black lines show the paths along which we
chose to take brightness profiles for our diagnostics.
\label{fig:pera}}
\end{figure}

Figure\,\ref{fig:pera} shows a 50\,ksec {\em ROSAT} HRI exposure of Perseus
A\footnote{These data were obtained from the LEGACY public archive situated
at GSFC (NASA).}.  Per A is a radio galaxy with an estimated kinetic power
of $\gtrsim 10^{43}{\rm\,ergs\,sec^{-1}}$ (Pedlar et al.~1990), at a
redshift of $z\approx 0.02$ or $80\,{\rm Mpc\,{h_{\rm 75}}^{-1}}$.  It is
located in a dense cluster environment with core densities of $n_{\rm
0}\approx 0.02-0.1\,{\rm cm^{-3}}$, a core radius of $r_{\rm c} \gtrsim 50$
kpc, and a temperature of $\approx 7$ keV (see White \& Sarazin 1988).  The
elliptical shell structure is readily visible from the plot and has been
the subject of a paper by B\"ohringer et al.~(1993).  The shell semi--minor
and semi--major axes are approximately $30''$ and $45''$, respectively,
corresponding to $12$ and $17$ kpc for a Hubble constant of $H=75\,{\rm
km\,sec^{-1}\,Mpc^{-1}}$.  Because the shell is so well--defined over a
significant angle, we are confident that the source is still in supersonic
expansion or has only recently crossed the sound barrier.  It is also
obvious from the image that our assumption of a spherically symmetric,
stationary cluster medium is idealized --- the bright feature to the east
indicates that ``cluster weather'' has probably had a significant impact on
the appearance of the structures.  However, the brightness changes are
moderate and, keeping in mind those caveats, we feel justified in applying
our model to this source with some caution.

\begin{figure}
\centerline{
  \psfig{figure=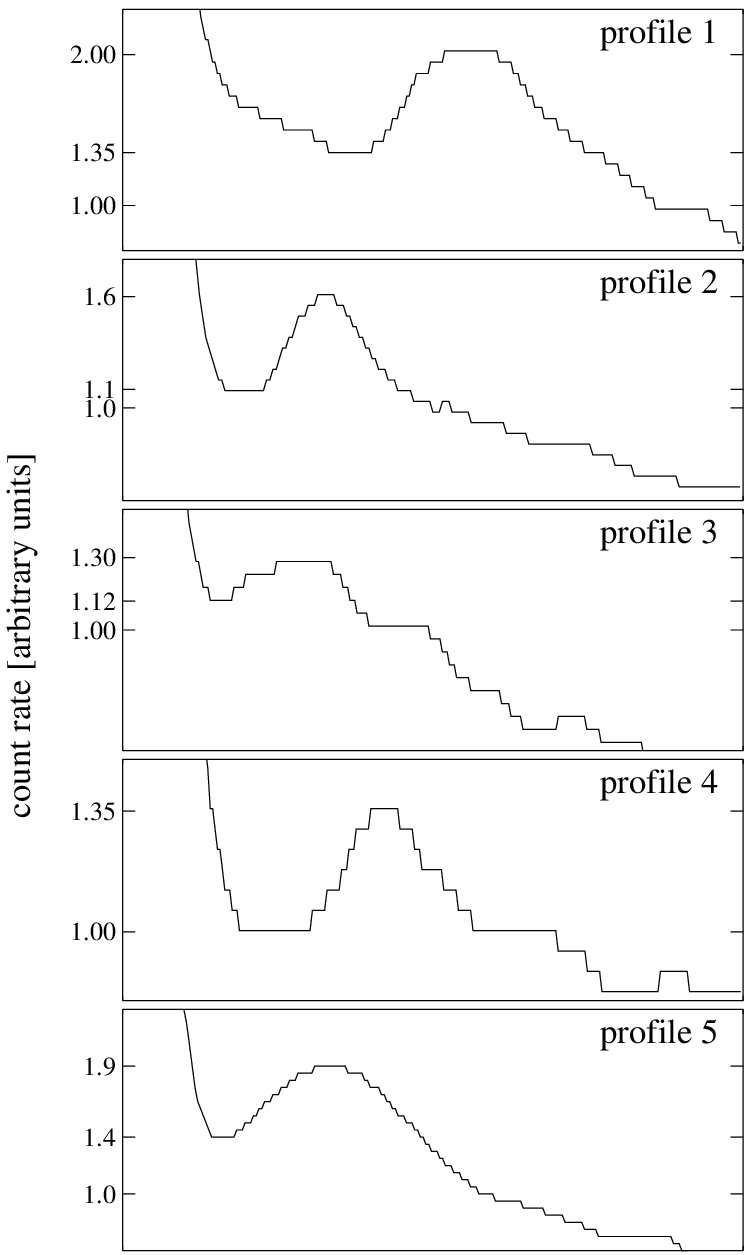,width=.5\columnwidth}}
\caption{Brightness profiles 1 through 5 according to
Fig.\,\ref{fig:pera} in arbitrary units.\label{fig:profiles}}
\end{figure}

\begin{figure}
\centerline{
  \psfig{figure=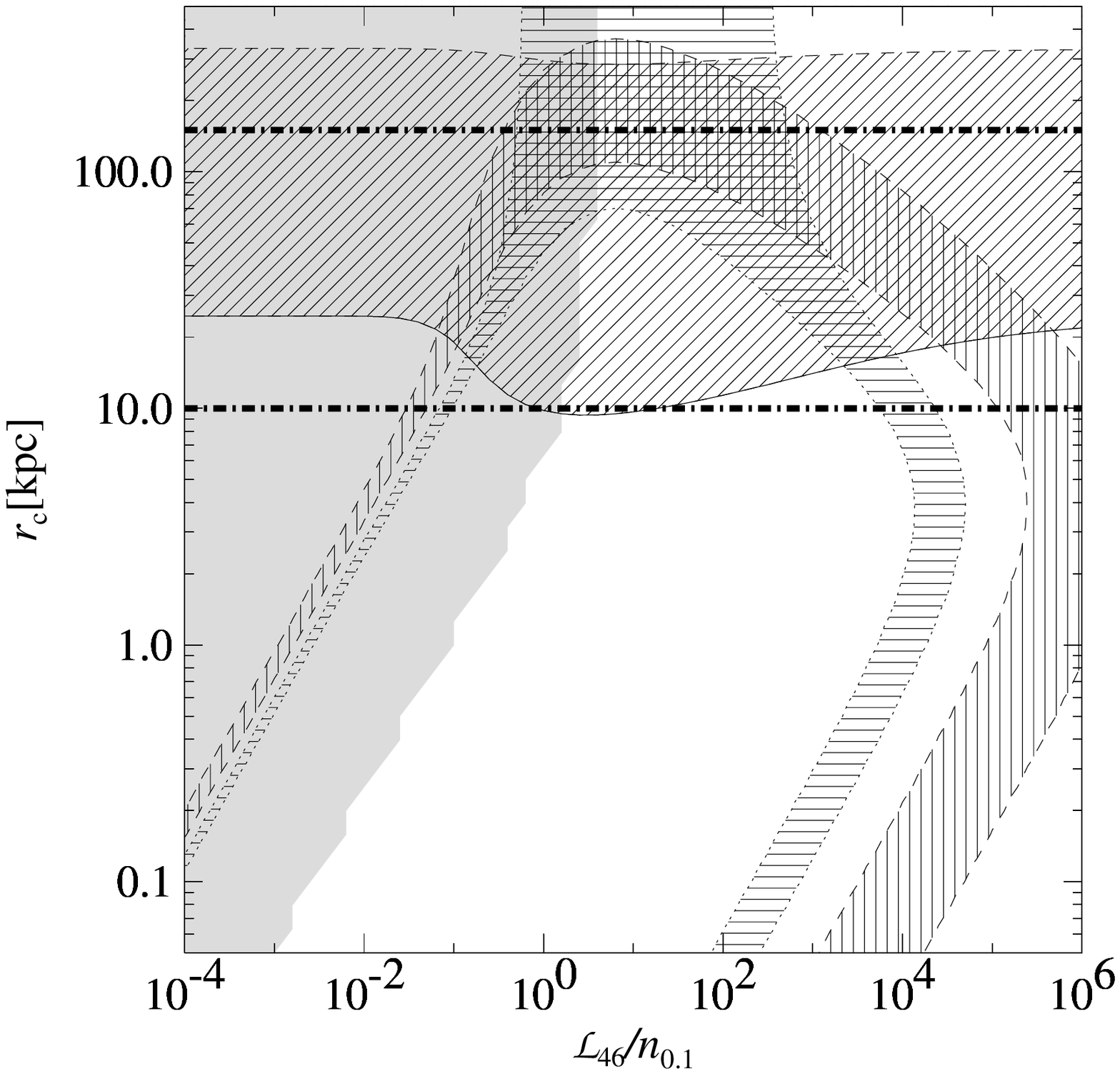,width=.5\columnwidth}}
\caption{Model contours in parameter space for Perseus A,
assuming isothermal cluster gas at $7$\,keV and a shell size of 16\,kpc,
seen through the {\em ROSAT} HRI response.  The three hatched regions
correspond to the observational diagnostics described in \S
\ref{sec:diagnostic}: a) the observed {\em shell-cluster ratio} of $\approx
2$ (hatched vertically), b) the observed {\em center-cluster ratio} of
$\approx 1.1$ (hatched horizontally), and c) the limits set by the observed
{\em shell count rate} of $\gtrsim 0.64\,{\rm counts\,sec^{-1}}$ and the
estimated central density of $n_{\rm 0}\lesssim 0.1\,{\rm cm^{-3}}$
(Sarazin \& White 1989, hatched diagonally).  Keep in mind that the
predicted count rates for a given $\scrL/{\rho_{\rm 0}}$ and $r_{\rm c}$
still scale with an additional ${n_{\rm 0}}^{2}$, thus the count rate only
allows us to set limits in this plot.  The grey area shows the subsonic
region in parameter space. The two thick dash--dotted lines correspond to
the estimated core radius of Per A (White et al.~1997, White \& Sarazin
1989).
\label{fig:rosat}}
\end{figure}

We have computed our model for a grid of various $r_{\rm c}$ and $\scrL$.
The cluster gas was assumed to have a temperature of $kT\sim\,7$\,keV and a
central density of $n_0=0.1\,{\rm cm^{-3}}$.  The integration was stopped
at a size of 16\,kpc, the approximate size of the source.  We computed the
three diagnostics described in \S \ref{sec:diagnostic} assuming the {\em
ROSAT} passband.  We also calculated the region in $r_{\rm c}$--$\scrL$
space for which the source is still supersonic.  We find that for a core
radius $r_{\rm c}\gtrsim 50$\,kpc, ${\mathcal L}/n_{\rm 0}$ must exceed
$5\times 10^{46}\,{\rm ergs\,cm^{3}\,sec^{-1}}$, and for a density of
$n_{\rm 0}\approx 0.02\,{\rm cm}^{-3}$, the mean kinetic luminosity must be
$\scrL> 10^{45}\,{\rm ergs\,sec^{-1}}$ to satisfy the supersonic condition.
The shaded region in Fig.\,\ref{fig:rosat} shows the region in parameter
space which is forbidden if we insist that the source be supersonic.
 
To compare these models to the data for Perseus A, we took radial
brightness profiles at several selected locations (shown in
Fig.\,\ref{fig:pera} and Fig.\,\ref{fig:profiles}).  We decided to
hand-pick these locations rather than assign them randomly due to the
complications caused by the ``cluster weather'' we pointed out earlier.
The enhancement at the shell compared to the brightness just outside ({\em
i.e.}, the {\em shell-cluster ratio}) is roughly\footnote{The HRI
resolution is of order 4 arcsec, which is roughly the expected width of the
shell at a radius of 40 arcsec (see equation \ref{eq:lambda}), thus we
expect the shell to be just marginally resolved.  The brightness ratios we
extract from the image will thus be lower limits.  The core is the dominant
feature and will have to be removed.} a factor of $1.3-2$.  The region
corresponding to these values in parameter space is shown in
Fig.\,\ref{fig:rosat} as a vertically hatched area with a dashed border.
Approximating the core--subtracted central surface brightness by the
brightness minimum between core and shell, we find a brightness ratio of
the interior against the cluster immediately outside the shell (i.e.,
the {\em center-cluster ratio}) of $\approx 1-1.3$.  The region in $r_{\rm
c}$--$\scrL\over\rho_{\rm 0}$ space compatible with this condition is also
shown as a hatched area (horizontal lines) with a dotted border in
Fig.\,\ref{fig:rosat}.  We also estimated the total count rate for the
shell area (i.e., the {\em shell count rate $\zeta$} from \S
\ref{sec:diagnostic}) to be $\zeta\approx 0.64\,{\rm s^{-1}}$ from taking
an elliptical ring aperture.  To display the predicted total count rate in
the same plot we will have to remove a factor of $(\rho_{\rm 0}/d)^{2}$
from this value (see section \ref{sec:diagnostic}).  Using the assumed
density from Sarazin \& White (1988) of $\rho_{\rm 0}\approx 0.02 - 0.1\,
{\rm cm^{-3}}$ and $d=80$\,Mpc, we can calculate the corresponding diagonally
hatched area in Fig.\,\ref{fig:rosat}.

White \& Sarazin (1988) provide an estimate for the core radius of $r_{\rm
c}\approx 10$\,kpc, whereas White, Jones, \& Forman (1997) chose a core
radius of 150\,kpc.  These values for $r_{\rm c}$ are shown as two thick,
dash--dotted lines in Fig.\,\ref{fig:rosat}.  Together with the other
constraints in Fig.\,\ref{fig:rosat}, these estimates allow solutions in a
range of ${\mathcal L}_{\rm 46}/n_{\rm 0.1}$ from $\approx 5\times 10^{-1}$
to $5\times 10^{2}$.  The lower end of this range can be ruled out by the
requirement of supersonic expansion.  Keeping in mind the rather large
uncertainty introduced by our measurements and in the input parameters we
used we find that the different areas in Fig.\,\ref{fig:rosat} match up in
a self--consistent fashion. 

The fact that the {\em average} kinetic luminosity is $\scrL > {\rm few}
\times\,10^{45}\,{\rm ergs\,sec^{-1}}$ is in itself a very interesting
result, as it suggests that the total power output of Per A is
significantly higher than the simple estimates based on the equipartition
energy content of the cocoons (Pedlar et al.~1990).  A possible conclusion
might be that equipartition is not a good approximation in this case (the
particle pressure will most likely exceed the magnetic pressure).  It is
interesting to note that a kinetic power estimate based on the conversion
factors by Bicknell, Dopita, \& O'Dea (1997) suggests that the {\em
instantaneous} kinetic luminosity is $L \lesssim 10^{44}\,{\rm
ergs\,sec^{-1}}$.  Hence this may be evidence that Perseus A is in a
relatively quiescent state (maybe corresponding to the ``{\em off}'' state
of Reynolds \& Begelman 1997). 

\subsection{Other Examples}
\label{sec:others}

Another well known source which exhibits an X--ray structure similar to the
one we propose here is Cygnus A.  This source has been object of an
extensive study by Clarke et al.~(1996), who ran a 3--dimensional
simulation of a jet advancing into a surrounding cluster medium.  Even
though there are cavities and brightness excesses visible at the center and
the rim of the cocoon, respectively, the structure is not nearly as
reminiscent of a shocked shell as the one in Per A.  It is more than twice
as distant as Perseus A and also exhibits a much stronger cooling flow,
both of which will tend to make the analysis we suggest harder.  We
therefore decided not to apply our method to this source, although this
would be an excellent target for application of our model once future high
resolution data are available. 

There are a number of other sources for which suspicious holes in the
X--ray morphology have been detected, for example in A4059.  However, the
data quality for those observations is not satisfactory yet and we will
have to postpone a closer analysis of those sources to future work. 

\section{Application to Future ({\em AXAF}) Data}
\label{sec:predictions}

\subsection{Detectability of Young Sources}
\label{sec:young}
Young sources are generally hard to detect in all wavelength bands, because
they are small and evolve quickly, so we expect the local density of young
sources to be very small, whereas distant sources will be faint and
unresolved.  They will be particularly hard to detect if they are in an
``off-state'' (in the sense of Reynolds \& Begelman 1997), in which case
radio surveys will most likely select against them, since the radio hot
spots fade away quickly and the spectrum becomes very steep.  In such a
case, optical observations of shock-excited H$\alpha$ emission, and the
X--ray emission that we are investigating here could provide ways of
finding such candidates.  The main issue with these sources is
detectability. 

The best spatial resolution we can hope for in X--ray imagers in the near
future is of order 0.5 ({\em AXAF} HRC) to 1\,arcsec ({\em AXAF}
ACIS). Thus, in order to resolve a source, it has to subtend more than
2\,arcsec on the sky (this assumes that the core, which is by far the
dominant feature, will not contaminate more than the central resolution
element).  But even if a source can be resolved, we shall show that
extremely long observing times will be necessary to achieve significant
signal--to--noise: even though the surface brightness is higher for small
sources, the cluster brightness also rises toward the center. 

\begin{figure}
\centerline{
  \psfig{figure=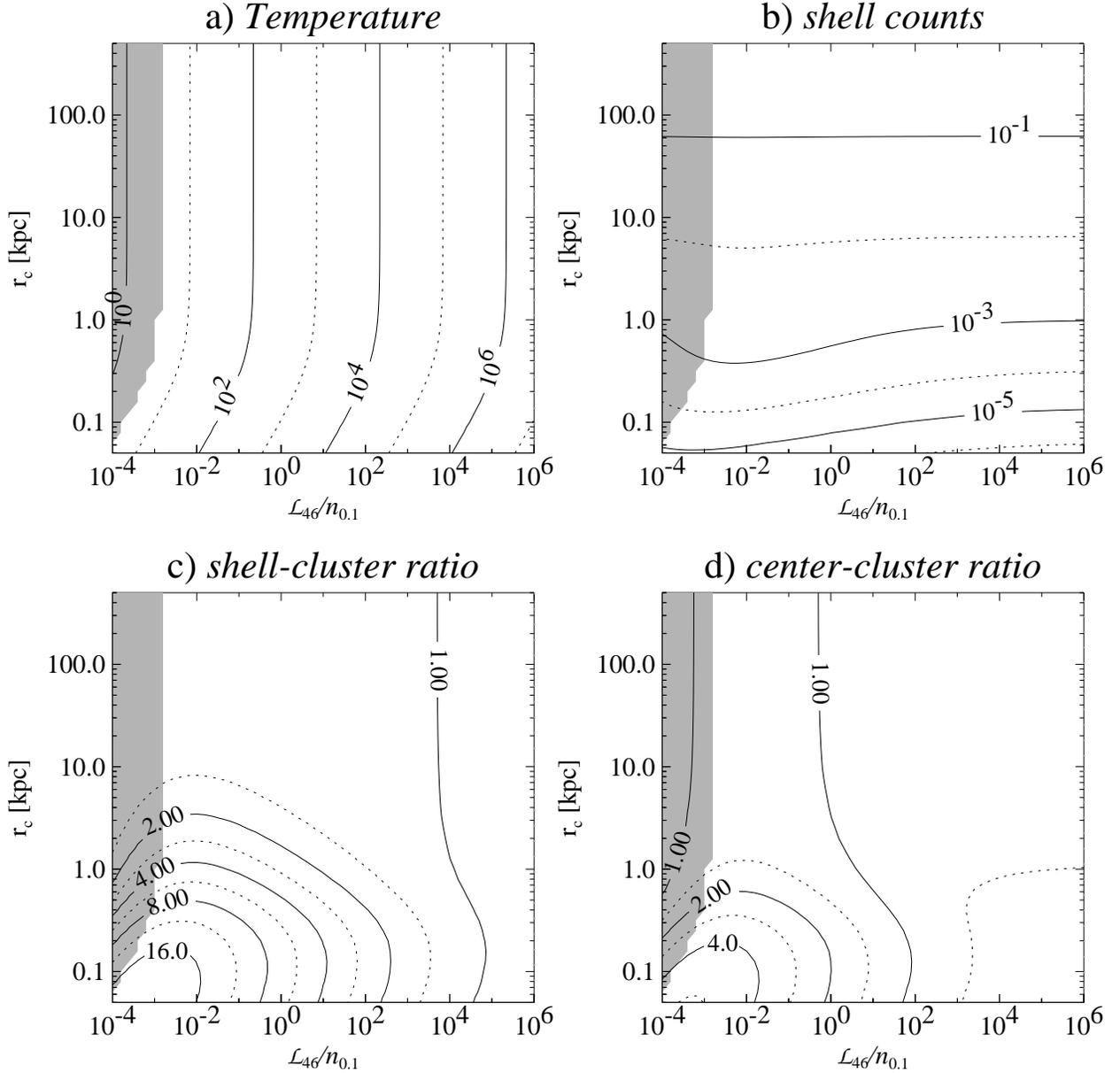,width=\textwidth}}
\caption{Contour plots of the shell temperature and our three
diagnostics (see \S \ref{sec:diagnostic}) as functions of the source
parameters $r_{\rm c}$ and $\scrL/\rho_{\rm 0}$ for a source with shell
radius 500 pc at a distance of 100\,Mpc.  The contour levels are separated
by factors of 10 in plots a) and b) and by factors of $\sqrt{2}$ in plots
c) and d). We assumed the {\em AXAF} ACIS response.  The grey area on the
left side of the plot indicates parameter values for which a source of this
size has turned subsonic for a cluster temperature of 4 keV.  The following
quantities are shown: a) Temperature in keV, b) {\em shell count rate
$\zeta$}, normalized to $n_{\rm 0}=0.1\,{\rm cm^{-3}}$, i.e.,
$\zeta\,\left(0.1\,{\rm cm^{-3}}/n_{\rm 0}\right)^2$. At large core radii,
cluster emission dominates, producing horizontal lines in the figure. c)
{\em shell-cluster ratio}, d) {\em center-cluster ratio}.
\label{fig:small}}
\end{figure}

We have calculated the three observational diagnostics from \S
\ref{sec:diagnostic} for a source of size 500\,pc over a grid of 2100
parameter combinations of $r_{\rm c}$ between 50\,pc and 500\,kpc and
$\scrL$ between $10^{42}$ to $10^{52}\,{\rm ergs\,sec^{-1}}$, assuming the
predicted {\em AXAF} ACIS passband.  The result is shown in
Fig.\,\ref{fig:small}. 

The shell temperature is generally higher for small sources.
Fig.\,\ref{fig:small}a shows the temperature as a function of ${\mathcal
L}/\rho_{\rm 0}$ and $r_{\rm c}$ for a shell radius of 500\,pc.  The
hatched region shows the parameter values for which a source of this size
at the fiducial cluster temperature of 4\,keV has become subsonic.  Even at
such a young age, we see that low power sources or sources in a very dense
environment have already stalled.  For higher luminosities the solution is
well behaved and basically follows the scaling relations
(\ref{eq:selfsimilar}) - (\ref{eq:scaleconstant}).  For very high
luminosities/low densities the temperature formally reaches values in
excess of 100 keV, beyond which our model is certainly not valid, as we
assumed the shell gas to be strictly non--relativistic. 

Figure\,\ref{fig:small}b shows the calculated {\em shell count rate $\zeta$}
for a source at a distance of 100 Mpc, normalized to a central density of
$n_{\rm 0}=0.1\,{\rm cm^{-3}}$, so what is plotted is
$\zeta\times\left(\frac{0.1\,{\rm cm^{-3}}}{n_{\rm 0}}\right)^2$.  Since
$\zeta$ increases with $n_{\rm 0}^2$, sources in dense cluster environments
will be much easier to detect.  The figure shows that for big and dense
clusters we might hope to detect a sufficient number of the photons from
the shell. 

In addition to a high count rate a second issue for detectability is the
contrast of the shell structure (i.e., the {\em shell-cluster
ratio}).  Fig.\,\ref{fig:small}c shows that the {\em shell-cluster ratio}
is significant over a large range of luminosities and core radii.  However,
for sources still within the core of the cluster profile the ratio drops
and approaches one (i.e. undetectability), since in such cases the
line of sight passes through a deep for- and background screen of cluster
gas, which tends to swamp out the shell emission.  Thus we have a detection
dilemma: For sources with large core radii, the count rates are high but
the contrast is small.  For sources which have broken out of the flat part
of the cluster gas, the contrast is high, but the count rates drop.  There
is an optimal range in core radius and luminosity, which is the larger, the
higher the density of the host environment is. 

Figure\,\ref{fig:small}d shows the {\em center-cluster ratio} to indicate
whether we see the cocoon as an X--ray hole or whether it shows a
brightness enhancement relative to the cluster gas.  In the presence of a
bright core, it will presumably be easier to detect a source which shows a
central brightness depression against the cluster, since in such a case we
will see a dark ring between core and cluster. 

\begin{figure}
\centerline{
  \psfig{figure=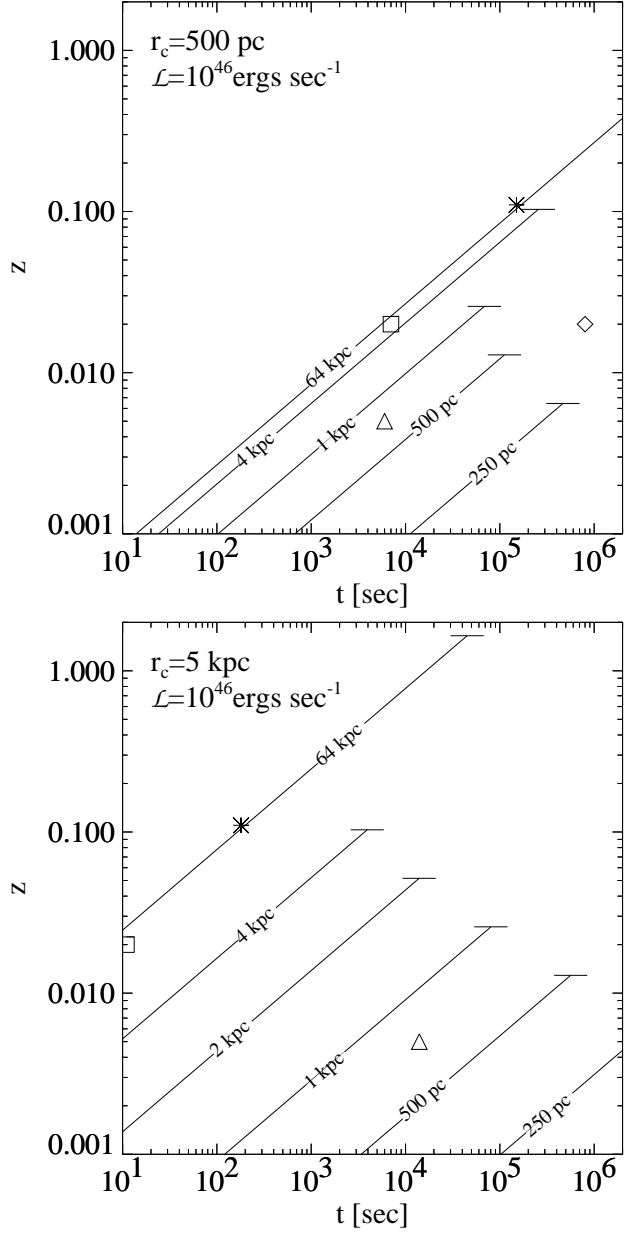,width=.5\columnwidth}}
\caption{Detectability constraints. This figure shows the
exposure times needed to measure $\delta$, the difference in the count rate
from the annulus between $r_{\rm c}$ and $r_{\rm s}$ to the count rate from
an annulus of equal area lying just outside $r_{\rm s}$, to within 30 per
cent.  This demonstrates the non-zero value of this difference to 3 sigma
confidence.  The lines correspond to different source sizes as indicated in
the figure. They are terminated when the spatial resolution limit of {\em
AXAF} ACIS is reached, i.e., when the source becomes smaller than 4 arcsec
across. Since it is properly oversampled, the {\em AXAF} HRC has twice the
resolution, but a lower effective area and no energy resolution.  The
symbols correspond to different sources we picked to demonstrate what can
be achieved. Star: VII Zw485, square: Per A, diamond: 4C34.09, triangle:
NGC1052.
\label{fig:detectability}}
\end{figure}

We conclude that the redshift constraints and the fact that bright cores
will tend to render the shells of barely resolved sources invisible ({\em
i.e.}, the core will swamp out the shell) poses a serious problem for the
detectability of young sources.  Generally, though, it should be possible
to detect close sources in sufficiently dense environments.  The
detectability constraints have been compiled in Fig.\,
\ref{fig:detectability}.  This figure has been constructed as follows.  For
a given set of source parameters and distance, we calculate the count rate
in the `observed' shell (i.e.  the annulus between $r_{\rm c}$ and $r_{\rm
s}$), $C_{\rm shell}$.  We also compute the count rate in an annulus of the
same area lying just outside the observed shell, $C_{\rm cluster}$.  The
shell is deemed to have been detected if the brightness profile is seen to
possess a jump at $r_{\rm s}$.  Thus, we calculate the exposure time needed
to measure $\delta=C_{\rm shell} - C_{\rm cluster}$ to within 30 per cent
(i.e. the exposure time required to demonstrate the non-zero value of delta
to 3 sigma confidence).  We have used $\scrL_{\rm 46}/n_{\rm 0.1}=1$ and
$r_{\rm c}=500\,$pc and $r_{\rm c}=5\,$kpc.  In terms of detectability this
is a more meaningful quantity than the actual brightness ratio of the two
rings, since a source is only detectable if there is a visible jump in
brightness across the shell.

\subsection{Extended sources}
\label{sec:old}
As was explicitly shown in \S \ref{sec:PerA} for Perseus A, application of
our observational diagnostics to a well resolved source can place
constraints on the source parameters, especially if the core radius of the
cluster is known. 

\begin{figure}
\centerline{
  \psfig{figure=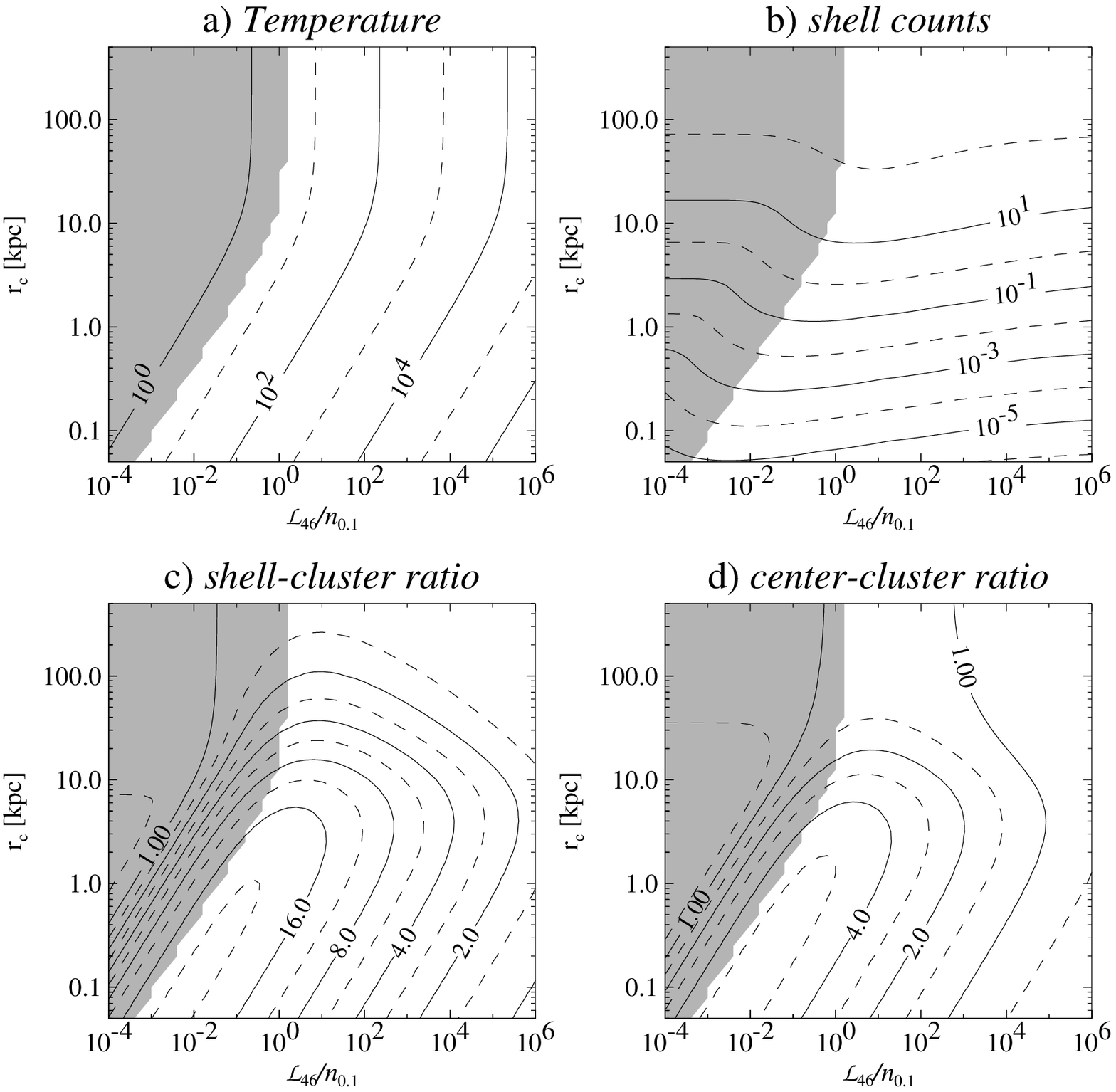,width=\columnwidth}}
\caption{Same as Fig.\,\ref{fig:small} for a
source of 16 kpc radius. \label{fig:big}}
\end{figure}

We have computed our three observational diagnostics from \S
\ref{sec:diagnostic} for a much more extended source of size 16 kpc, as
shown in Fig.\,\ref{fig:big}.  At this size, only powerful sources will
have maintained a supersonic coasting speed, and we can expect the
transition between super-- and subsonic sources to be a strong indicator of
the overall source power, since the break in the data happens at a well
defined value of $\scrL/\rho_{\rm 0}$ for given core radius $r_{\rm c}$ and
cluster temperature $T_{\rm cluster}$.  The temperature shows the
dependence on core radius and power we expect from the scaling relations in
equations (\ref{eq:selfsimilar})-(\ref{eq:scaleconstant}). 

Because large sources (by definition) subtend a larger solid angle, their
count rates can be much higher than those for small sources, even though
the peak shell surface brightness decreases.  Since we include the fore--
and background cluster emission in the {\em shell count rate} (shown in
Fig.\,\ref{fig:big}b), this effect becomes even more pronounced.  Also,
since the shell will likely be resolved and separated from the core, the
recognition of the structure itself and the reduction of shell parameters
will be simpler than for small sources.  The strong dependence of the {\em
shell count rate} on the core radius for $r_{\rm c}\lesssim16$\,kpc is
introduced because we fixed the central density, thus the mass enclosed in
a sphere radius $r_{\rm s}$ (i.e., the swept up mass) increases with
the core radius, and so does the shell density.  For $r_{\rm
c}\gtrsim16$\,kpc the dependence is not as strong.  In this case, it is
produced by the fact that we integrate over a longer line of sight of
undisturbed cluster gas within the core for larger $r_{\rm c}$.  As in the
case of Perseus A, we can see that the total count rates are in a
comfortable regime to achieve good statistics over a wide parameter range
for a cluster with our fiducial density. 

If the core radius can be measured by other means (e.g., from optical or
X-ray data) and if the redshift of the source is known, the brightness
ratios in Fig.\,\ref{fig:big}c and \ref{fig:big}d can be used to constrain
$\scrL/\rho_{\rm 0}$ to no more than two possible values in the
$\scrL/\rho_{\rm 0}$--$r_{\rm c}$--plane, independently from the total
count rate (if the supersonic condition is applicable we can fully
constrain this parameter in many cases).  In this case, the total count
rate can be used to determine $\rho_{\rm 0}$ and $\mathcal L$ separately,
and thus measure the age of the source.  Density, age, and in particular
the kinetic luminosity will be useful input into jet models. 

Since some modern X--ray detectors will have high spectral as well as
spatial resolution (e.g., {\em AXAF} ACIS), it will be feasible to obtain
imaging spectra from resolved sources.  In the case of ACIS with a spatial
resolution of about 1 arcsec, the source has to be larger than 2 arcsec in
radius, thus the redshift restriction will be $z<h_{\rm 75}/8$.  In the
regime we are interested in the bremsstrahlung emissivity goes as
$T^{-1/2}e^{-h\nu/kT}\ln{\left[kT/(h\nu)\right]}$, and since the shell temperatXure is
mainly dependent on $\scrL/\rho_{\rm 0}$, spectra are very useful to
separate out this parameter.  A color--color--image would emphasize regions
of different temperature (i.e., the shell versus the undisturbed
X-ray atmosphere).  For high enough count rates it will even be possible to
obtain information about the foreground absorption.  Spectra can also give
a handle on the composition of the cluster gas from spectral line analysis.


\section{Discussion}
\label{sec:discussion}

\subsection{Complications}
The model we have discussed in the previous sections offers a simple way of
probing the radio source structure and determining important
parameters. However, since it is a gross oversimplification of reality, we
have to discuss several complications. 

\subsubsection{Intermittency}
\label{sec:intermittency}

As recently proposed by Reynolds \& Begelman (1997), a large population of
sources might be intermittent (i.e., strongly time variable in their
kinetic energy output), thereby explaining the observed size distribution
of CSOs and MSOs, which shows a previously not understood flattening at
small sizes.  This could be due to the fact that sources spend a
significant fraction of their lifetime in a quiescent state, in which the
jets are turned ``{\em off}''.  In such a case, the shell and the cocoon
are still expanding, but they will slow down and the shell will thicken.
Since a large part of the radio luminosity comes from the radio hot spots,
which fade away rapidly, the radio luminosity also decreases.  Whereas for
large sources the cocoons filled with relativistic plasma provide enough
emission to be detected in the radio, young dormant sources will be very
hard to detect with radio observations.  Thus we might hope to detect the
X--ray signatures of such sources. 

The major difference that has to be incorporated into our analysis is a
time variable $L(t)$ instead of $\mathcal L$ in equation (\ref{eq:cocoon}).
Since we integrated the equations numerically, this change presents no
difficulty.  For simplicity we followed Reynolds \& Begelman (1997) in
using a ``picket-fence function'' for $L$.  We chose a duty cycle of 10\%,
i.e., the source is ``{\em on}'' for 10,000\,years and ``{\em off}''
for 90,000\,years.  Depending on the average source power, this happens at
different evolutionary stages of the source, thus we should expect to see
changes from the previous figures.  However, as the source grows, the
influence of the intermittency becomes smaller and the behavior approaches
the solution for constant luminosity, corresponding to an average power of
$\scrL=\langle L(t)\rangle$.  For large sources, the results we presented
above are therefore essentially unaltered. 

\subsubsection{Cold Material and Mixing}
\label{sec:cold}

An important unknown we have neglected to include in our treatment is the
possible multi--phase nature of the host ISM/ICM.  A cold phase would be
hard to detect in an X--ray observation, although it can sometimes be seen
via X-ray absorption (Allen et al.~1995, Fabian 1994).  Spectra could help
in finding multiple temperature components in the continuum emission and
abundances of low ionization states from spectral lines and edges.  The
presence of a cold component could severely alter the dynamics of our
expanding shell if both filling factor and mass residing in the cold phase
were high enough. 

The general picture is that, as the shell reaches the cold blobs of
material, a shock is driven into the cold matter with the same strength
(i.e., same Mach number) as the shock into the ISM/ICM, since the
cold and hot phases are assumed to be in pressure equilibrium.  The cold
material will therefore be heated and radiate either in the X--ray regime
(if it is hot enough) or in the UV/optical.  Depending on their sizes, the
blobs could either be completely evaporated or their remnants might remain
inside the cocoon.  It might get shredded by hydrodynamical instabilities,
in which case it would mix with the shocked gas and the cocoon plasma.  The
cold material could cool very efficiently and might radiate away a lot of
the shell energy.  Future work will concentrate on the optical line
emission we would expect to see from such filaments (see, e.g., Bicknell \&
Begelman 1996). 

A related question is the possibility that dynamical instabilities (e.g.,
Rayleigh-Taylor, Kelvin-Helmholtz) could mix material from the shell and
the cocoon, producing pockets of non--relativistic material in the
relativistic cocoon.  If mixing is strong, a curtain of intermediate
temperature material could form and absorb radio emission from the cocoon,
thus producing the characteristic spectral shape of Gigahertz Peaked
Sources (GPS), as described by Bicknell et al.\,(1997). 

Foreground absorption might also affect our observational diagnostics. Even
though the column density of Galactic material is well known in most
directions, the presence of a cold component in the host cluster could have
a significant effect on the detected signal. As long as the covering factor
and the filling factor of the cold component are small, this effect should
be negligible. If, however, the cold matter covers a large cross section of
the source with sufficient column density, absorption can change the
spectral shape and therefore alter not only the total expected count rates
but also the brightness ratios, as they depend on the temperature
difference between the shell and the cluster.  Current data suggest that
typical values for the column density of $N_{\rm H}$ of cold material
intrinsic to such sources are of the order $N_{\rm H}\sim 10^{21}\,{\rm
cm^{-2}}$, too low to affect the {\em AXAF} band significantly. 

\subsubsection{Density Profile}
\label{sec:king}
Clearly, an isothermal King profile is a gross oversimplification of
reality, but it should at least provide us with a first order
approximation.  Introducing two separate King profiles (one for the
cluster, one for the Galaxy itself) might provide a better description of
reality, but it would also increase the number of free parameters by three
$(r_{\rm c}, \rho_{\rm 0}$, and $\beta)$, all of which would have to be
determined by other means to improve this analysis.  We decided that the
increase in realism would not justify the necessary computing time and the
uncertainty as to which values we should chose for the new parameters.

\subsection{Conclusions}
\label{sec:conclusions}
We have presented a very simple model of the evolution of powerful radio
galaxies into a surrounding ISM/ICM in order to make predictions about the
detectability and appearance of such sources for future X--ray missions.
We assumed uniform pressure, spherical symmetry, and a King profile density
distribution in the ambient medium to describe the cocoon and the shell of
swept up material and have calculated a grid of models for various source
parameters to provide observational diagnostics for high resolution X--ray
observations.

We are particularly interested in young sources, since signatures of
intermittency are most pronounced in the early stages of source
evolution. However, there is only a rather limited range of parameters for
which we might hope even to detect the shell for such sources, since the
expected count rates are low for all but the most dense environments. Also,
contamination by a bright core could make such detections impossible.

Larger sources offer more chances not only for detection but also for
analysis and application of our model grid.  For a source with known
redshift and core radius, it should be possible, at least in principle, to
determine the central density of the King profile, the average kinetic
source luminosity, and the source age.  The knowledge of these parameters
could help a great deal in understanding the process of jet formation.  As
an example we apply our model to a 50 ksec {\em ROSAT} HRI observation of
Perseus A, and find that the time averaged power most likely exceeds
$10^{45}\,{\rm ergs\,sec^{-1}}$.

\acknowledgements

This work has been supported by NSF grant AST-9529175.

\end{document}